\documentclass[psfig,useAMS,usenatbib]{mn2e}
\usepackage{times}
\usepackage{epsfig}
\usepackage{amsmath}
\usepackage{natbib}
\usepackage{longtable}

\def\apj {ApJ}

\def\apjs {ApJS}
\def\aj {AJ}
\def\aap {A\&A}
\def\mnras {MNRAS}
\def\pasp {PASP}
\def\zphot {z_{\rm phot}}
\def\zspec {z_{\rm spec}}

\title[Photometric redshifts and K-corrections for SDSS-DR7]
  {Photometric redshifts and K-corrections for Sloan Digital Sky Survey Seven Data Release}
\author[O'Mill et al.]
  {Ana Laura O'Mill,$^{1,2}$\thanks{E-mail: aomill@oac.uncor.edu}
  Fernanda Duplancic,$^{1,2}$ Diego Garc\'\i a Lambas,$^{1,2}$   
  \& Laerte Sodr\'e Jr$^{3}$ \\ 
  $^1$ Instituto de  Astronom\'\i a Te\'orica y Experimental, IATE, Observatorio Astron\'omico, Universidad Nacional de C\'ordoba,\\ 
Laprida 854, X5000BGR, C\'ordoba Argentina\\
  $^2$Consejo de Investigaciones Cient\'\i ficas y T\'ecnicas (CONICET),\\Avenida Rivadavia 
1917, C1033AAJ, Buenos Aires, Argentina\\
  $^3$ Departamento de Astronomia, Instituto de Astronomia, Geof\'\i sica
e Ci\^encias Atmosf\'ericas da USP,\\ Rua do Mat\~ao 1226, Cidade
Universit\'aria, 05508-090, S\~ao Paulo, Brazil} 
\date{Released 2010 Xxxxx XX}
\pagerange{\pageref{firstpage}--\pageref{lastpage}} \pubyear{2002}
\def\LaTeX{L\kern-.36em\raise.3ex\hbox{a}\kern-.15em
    T\kern-.1667em\lower.7ex\hbox{E}\kern-.125emX}

\begin{document}
\label{firstpage}
\maketitle
\begin{abstract}
We present a catalogue of galaxy photometric redshifts and k-corrections for the Sloan 
Digital Sky Survey Seven Data Release (SDSS-DR7), available on the World Wide Web. 
The photometric redshifts 
were estimated with an artificial neural network using five \textit{ugriz} bands, 
concentration indices and Petrosian radii in the $g$ and $r$ bands. We have explored 
our redshift estimates with different training set concluding that the best choice to 
improve redshift accuracy comprises the Main Galaxies Sample (MGS), the Luminous Red Galaxies, 
and galaxies of active galactic nuclei covering the redshift range $0<z\leq 0.3$. 
For the MGS, the photometric redshift estimates agree with the spectroscopic values within $rms=0.0227$. 
The derived distribution of photometric redshifts in the range $0<\zphot\leq0.6$ agrees well with the 
model predictions.\\
k-corrections were derived by calibration of the \texttt{k-correct\_v4.2} code results for the MGS with the 
reference frame ($z=0.1$) $(g-r)$ colours. We adopt a linear dependence of $k$ corrections on redshift 
and $(g-r)$ colours that provide suitable distributions of luminosity and colours for galaxies up to 
redshift $\zphot=0.6$ comparable to the results in the literature. Thus, our k-correction estimate procedure 
is a powerful, low computational time algorithm capable of reproducing suitable results that can be used 
for testing galaxy properties at intermediate redshifts using the large SDSS database.\\

\end{abstract}


\section{Introduction}

The knowledge of distances to galaxies is important to deduce intrinsic galaxy properties (absolute 
magnitude, size, etc.) from the observed properties (colours, sizes, angles, flux, apparent size, etc.).\\ 
Since pioneering works measuring redshift of bright galaxies
(e.g., (\citet{shapley1932}, \citet{humason1956}), many efforts have been invested 
in mapping the light and matter in the Universe. 
Statistical analysis of galaxy properties and systems can be invaluable
tools to study large-scale structure and evolution in the universe.\\
In recent years, multi-band photometry has been performed for several millions of
galaxies, whereas spectroscopic redshifts have been measured only
for a small fraction of the photometric data. The Sloan Digital Sky survey has obtained multi-band images 
for approximately one hundred billion galaxies (and the next surveys 
foresee increasing the number of objects up to billions), while spectroscopic measurements have 
been obtained for nearly one million galaxies. A solution to the difficulty of obtaining spectroscopic 
redshifts relies on the use of photometric redshift techniques. Although the redshifts calculated 
through these techniques are far less accurate than the spectroscopic measurements, 
these approximate distance estimates allow for useful analysis in fields such as extragalactic 
astronomy and observational cosmology.

Two basic family of methods are commonly employed to calculate photometric 
redshifts. In the template matching approach, a set of spectral energy 
distribution (SED) templates is fitted to the observations (e.g., colours).
In the empirical approach, on the other side, photometric redshifts are 
obtained from  a large and representative training set of galaxies with
both photometry and precise redshift estimations. The advantage of the 
first method is that it can also provide additional information, like 
the spectral type, k-corrections and absolute magnitudes. The accuracy 
of these estimations is limited by the SED models. The empirical model 
overcomes this limitation through the use of a training set which, however, 
should be large and representative enough to provide accurate
redshift estimations.\\

The observed spectral energy distribution of distant galaxies is redshifted 
with respect to that in the galaxy rest frame.
The k-correction term \citep{oke,hogg01} applied to the apparent magnitude 
measured in a given photometric band takes into account this effect,
allowing to compare photometric properties of galaxies at different
redshifts. The estimation of  k-corrections, then, is a requirement for many
studies of distant galaxies. It is possible to model k-corrections as a 
function of redshift and galaxy morphological type \citep{fuku95, Mannucci01}. 
\citet{lahav95} and \citet{banerji10} use ANNs to obtain morphological classification
of galaxies. However, these techniques employ as training sets, objects that had been 
classified by human eye subject to some degree of 
ambiguity and uncertainty, particularly at large redshifts. 

A more direct way to obtain k-corrections is by modelling 
galaxy SEDs as a function of wavelength. Usually template fitting of 
observed galaxy fluxes is employed to reconstruct the SED of the galaxy 
(\citet{blanton03}, \citet{blanton07}). The feasibility and accuracy of
this method relies in the quality of the models. 

For objects with spectral data, k-corrections can also be obtained directly. 
\citet{roche} used this technique to calculate k-corrections for early-type 
galaxies from the SDSS-DR6, providing individual estimates for each galaxy. 
However, this technique is restricted to a limited number of galaxies with 
spectroscopy.

Recent works (\citet{chilinga}, \citet{westra}) have
approximated k-corrections with analytical functions of redshift, 
parametrized with some property characterizing galaxy type.
\citet{chilinga} used different observed colour indices to approximate 
k-corrections for nine filters ($ugrizYJHK$). \citet{westra} used spectra 
from the Smithsonian Hectospec Lensing Survey to obtain direct measurements 
of k-corrections by parametrization with the ratio of the average flux red and 
bluewards of the 4000\AA{} break ($D_n4000$). These kinds of parametrization 
simplify the computation of k-corrections and improve their accuracy.\\
    
In this paper, we present a galaxy photometric redshift ($\zphot$) catalogue and a 
method for calculating k-corrections, for the seventh Data Release (DR7) 
of the Sloan Digital Sky Survey (SDSS) imaging catalogue (\citet{blanton03}, 
\citet{lrgs}, \citet{gunn98}, \citet{mgs}, \citet{sdss}). 
To compute photometric redshifts we used the ANNz software package 
(\citet{annz}), which have been shown to be a reliable tool.
There are also two sets of photometric redshifts in the SDSS database:
\citet{dr7} employ empirical, template-based 
and hybrid-techniques approaches to photometric redshift estimation, whereas 
\citet{zp} adopt a neural network method and provide two different estimations,
D1 and CC2. Here we also compare our redshift estimations with those in
the CC2 catalogue from \citet{zp}, which uses colours and 
concentration indices to infer redshifts. For the computation of 
k-corrections we propose a joint parametrization in terms of redshift 
and the $(g-r)$ colour in a certain reference frame  for all SDSS bands, 
as well as an algorithm to
determine them from the photometric data. 
We compare our results with those found in literature.

This paper is organized as follows. In Section 2 we describe the data used in 
our analysis. Section 3 presents our approach to calculate photometric 
redshifts in SDSS-DR7, analysing their advantages and limitations. 
Our estimation of k-corrections is presented in Section 4. Finally,
Section 5 summarizes the results obtained in this work.

Throughout this paper, we adopt a cosmological model characterized by
the parameters $\Omega_m=0.3$, $\Omega_{\Lambda}=0.7$ and $H_0=75~h~
{\rm km~s^{-1}~Mpc^{-1}}$.


\begin{table*}
\begin{minipage}{175mm}
\caption{Description of the samples used in this work.}
  \begin{center}\begin{tabular}{@{}ccl@{}}
  \hline
  \hline
sample name &  number of objects & Description  \\
 \hline
Sz1 & $\sim 550000$ & Selected from SDSS-DR7 spectroscopic data.\\ 
    &               & Used as training and validation sets in the computation of photometric redshifts \\
\hline
Sz2 & $\sim 70000$  & Selected from SDSS-DR7 MGS (excluding training set galaxies). \\
    &               & Used as testing set for photometric redshifts.\\
\hline
Sz3 & $\sim 82000$  & Selected from SDSS-DR6 photometric data.\\
    &               & Used  to compare photometric redshift estimation with \citet{zp}.\\
\hline
Sk1 & $\sim 122000$ & Selected from SDSS-DR7 MGS taking into account apparent magnitude and redshift limits (see text).\\
    &               & Used for k-correction calibration.\\
\hline
Sk2 & $\sim 575000$     & Selected from SDSS-DR7 photometric data with our photometric redshifts estimation. \\
    &               & Used to compute k-correction at higher redshift and compare with literature.\\
\hline
\hline
\label{t1}
\end{tabular}
\end{center}
\end{minipage}
\end{table*}

\section{The galaxy samples}

The samples of galaxies used in this work were drawn from the Sloan Digital 
Sky Survey Seven Data Release (SDSS-DR7, \citet{dr7}).
SDSS (\citet{sdss}) mapped more than one-quarter of the entire sky, 
performing  photometry and spectroscopy for galaxies, quasars and 
stars. SDSS-DR7 is the seventh major data release, corresponding to the
completion of the survey SDSS-II. It comprises $11.663$ sq. deg.
of imaging data, with an increment of $\sim2000$ sq. deg., over the 
previous data release, lying in regions of low Galactic latitude.
SDSS-DR7 provides imaging data for 357 million 
distinct objects in five bands, \textit{ugriz}, as well as
spectroscopy  over $\simeq \pi$ steradians in the North Galactic 
cap and $250$ square degrees in the South Galactic cap. 
The average wavelengths corresponding to the five broad bands 
 are $3551$, $4686$, $6165$, $7481$, and $8931$ \AA{} \citep{fuku96,hogg01,smit02}. 
For details regarding the SDSS camera see \citet{gunn98}; for astrometric 
calibrations see \citet{pier03}. 
The survey has completed spectroscopy over 9380 sq. deg.; the spectroscopy is now 
complete over a large contiguous area of the Northern Galactic Cap, closing the gap 
that was present in previous data releases. 

\begin{figure*}
\begin{picture}(450,240)
\put(0,0){\psfig{file=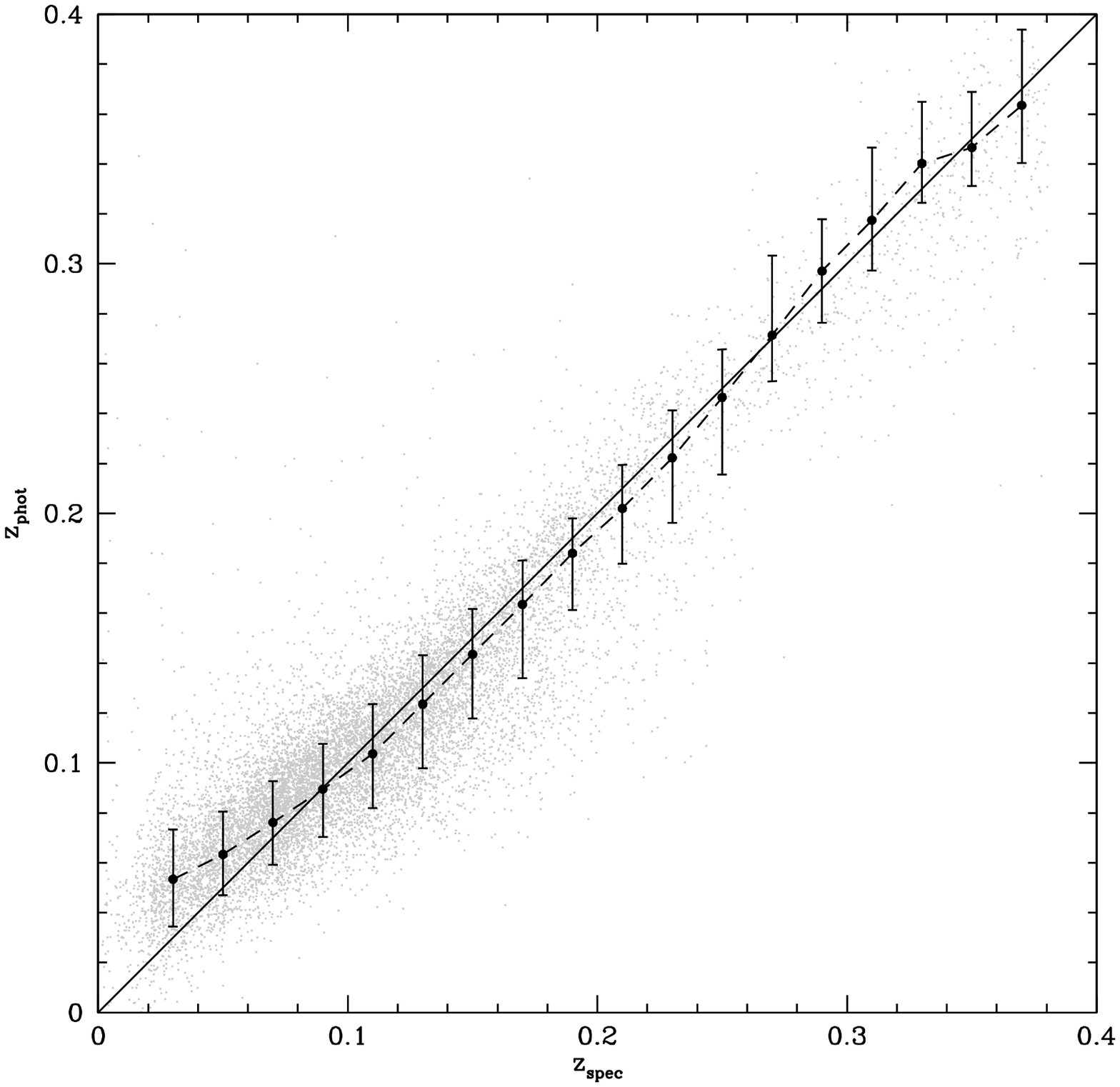,width=8cm}}
\put(240,0){\psfig{file=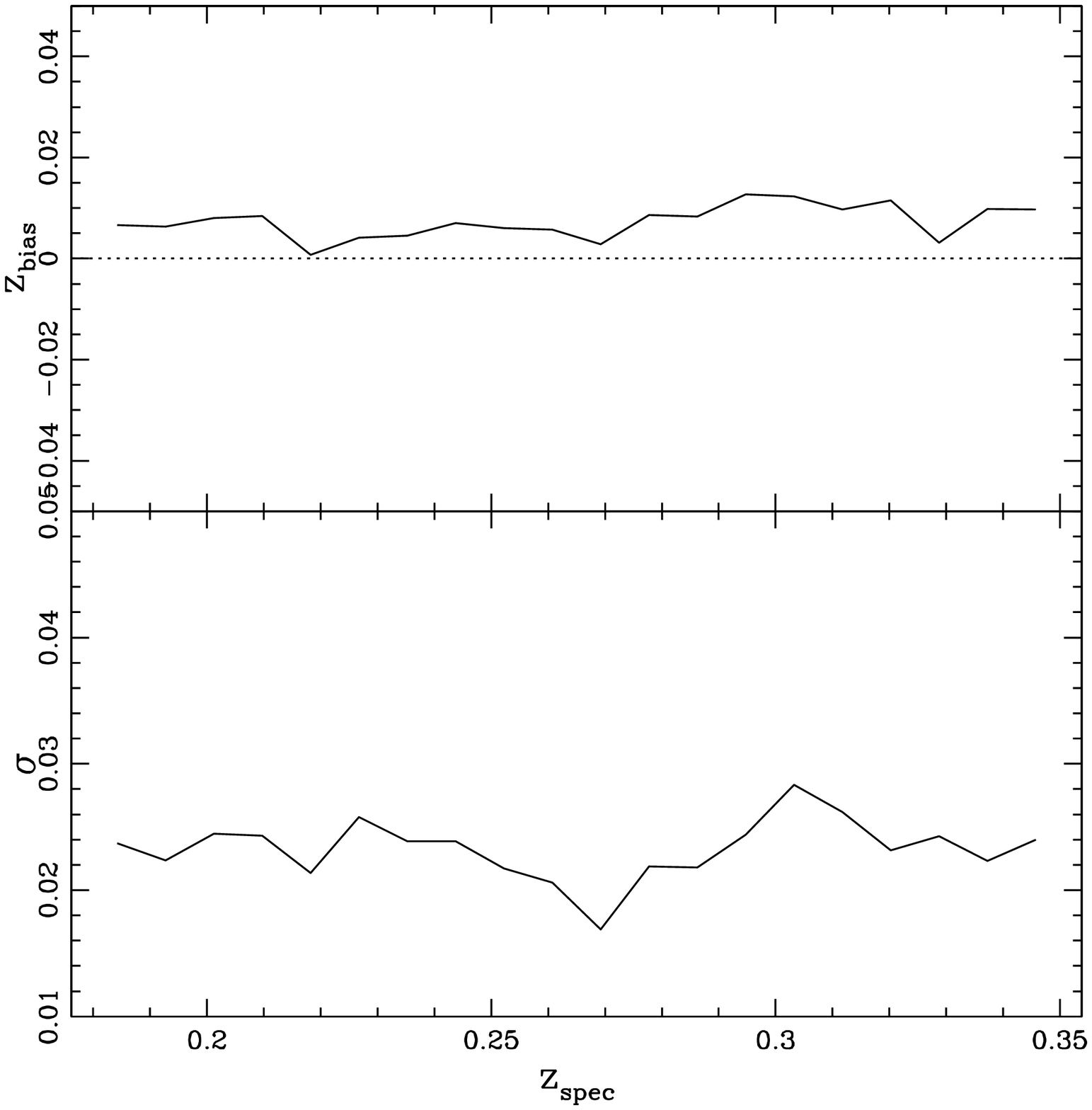,width=8cm}}
\end{picture}
\caption{Left panel: $\zspec$ vs $\zphot$ relation for Sz2 sample (grey dots). 
The dashed line corresponds to the median value of 
the photometric redshift estimated at a given spectroscopic redshift interval. 
Error bars correspond to the $10$ and $90$ percentiles associated to the median. 
The solid line shows the one-to-one relation. Right panels: systematic (top right)
and stochastic (bottom right) errors in the photometric redshift estimates.}
\label{f1}
\end{figure*}

In this work we have extracted two galaxy data sets from SDSS-DR7, one with spectroscopic redshifts measurements 
and the other consisting in photometric data.

The spectroscopic data was derived from the \texttt{fits} files 
at the SDSS home page\footnote{http://www.sdss.org/dr7/products/spectra/getspectra.html}. 
consisting in Main Galaxy Sample (MGS; \citet{mgs}), 
the Luminous Red Galaxy sample (LRG; \citet{lrgs}) and active galactic nuclei
(AGN; \citet{kauf2003}). These spectroscopic samples were used for the computation of photometric redshifts and calibration of k-corrections.

We built a first sample (hereafter Sz1) consisting in $80\%$ of the objects from MGS, 
$10\%$ from LRG and $10\%$ from AGNs, combined into a single set comprising $\sim 550000$ galaxies.
This sample was divided at random into two subsamples with the same number of 
objects generating a training and a validation set in order to calibrate the ANNz code used to 
the estimation of photometric redshifts.

As a testing set for the photometric redshifts we used a random sample (hereafter Sz2) of 60,000 objects from MGS. 
We excluded from this sample galaxies belonging to the training set 
to avoid undesirable biases.

For the calibration of k-corrections we used the full MGS. Following \citet{maglim}, we have taken 
the magnitude range where the number of galaxies per solid angle rises at a constant rate as a function of redshift 
in each SDSS band. Therefore, the apparent magnitude limits are set to ensure that the effect of incompleteness 
is small in our sample. The adopted apparent magnitude limits in 
each band are : $u<19.0$, $g<17.91$, $r<17.77$, $i<17.24$, and $z<16.97$. 
For these limits we consider galaxies with 
$z_{spec}<0.15$ to assure completeness in the $u$ and $g$ bands,
and $z_{spec}<0.18$ in the $r$, $i$ and $z$ bands. With these 
magnitude and redshift constraints, the sample used to calibrate the
k-corrections (hereafter Sk1) contains $\sim 122000$ galaxies.

The photometric data set were extracted from the \texttt{Galaxy} table of the \texttt{CasJobs}\footnote{http://cas.sdss.org/dr7/} database. 
We restrict our analysis to photometric objects with $r<21.5$, since this magnitude limit assures 
good photometric quality and a reliable star-galaxy separation (Stoughton et al., 2002, Scranton et al., 2002). 

In order to contrast our estimates with those obtained by similar methods, we compare our results 
with photometric redshifts from \citet{zp}. Taking into account these authors recommendation we 
use \texttt{PhotozCC2} estimates (obtained from colours and 
concentration indices, see \citet{zp}). It should be noted that \texttt{Photoz2} table was not 
updated for SDSS-DR7, so we have constructed a sample from SDSS-DR6 photometric data selecting at random 
82000 objects with \texttt{PhotozCC2} redshift information \footnote{SDSS-DR6 photometric data was downloaded from CasJobs 
just as DR7 photometric data including redshift information from Photoz2 table}. 
For this sample we calculated photometric redshifts through the methods described in this work. Hereafter 
we will call this sample as Sz3.

In order to compute k-correction at higher redshifts we have selected a random sample of $\sim 575000$ photometric SDSS-DR7
data for which we have determined photometric redshifts and k-corrections (sample Sk2).
For this sample we have also compared the k-corrections computed in this work 
with different results obtained from the literature.

Table  \ref{t1} summarizes the main characteristics of the different samples used in this work

\section{Photometric redshifts}

Photometric redshift techniques use photometric parameters
to perform an estimation of galaxy redshift.
This technique can be used to infer efficiently large numbers of galaxy 
distances, even for faint galaxies, for which spectroscopic measurements 
are prohibitive because they would require large amounts of telescope time. 

There are different techniques to estimate photometric redshifts
which can be classified into two groups. The first set of techniques makes use 
of a small number of model galaxy spectra derived from empirical or 
model-based  spectral energy distributions (SEDs). 
These methods estimate a galaxy redshift by finding an optimal combination 
of template spectra that reconstructs the observed galaxy colours
\citep[e.g.,][]{ben00,bolz00,csabai03}. The fact that these methods rely 
on a small number of template SEDs is their main disadvantage, in particular 
for galaxies at high redshifts, since a representative set of spectral 
templates applicable at all redshifts is not easy to obtain. 
The second group, called empirical methods 
\citep[e.g.,][]{con95,brun99}, comprises techniques that need a large amount 
of prior redshift information, in general in the form of training set.
This class of methods aims to derive a parametrization for the redshift as a 
function of photometric parameters. The form of this parametrization is 
obtained through the use of a suitably large and representative training set 
of galaxies for which we have both photometry and precisely known redshifts. 
In this case we can use combinations of galaxy photometric parameters,
such as magnitudes in different photometric bands, galaxy colours, and 
concentration indices. The main drawback of empirical methods is that the
training set should be representative of the sample of galaxies for which
we want to estimate photometric redshifts.\\

\begin{figure*}
\leavevmode \epsfysize=12cm \epsfxsize=9cm \epsfbox{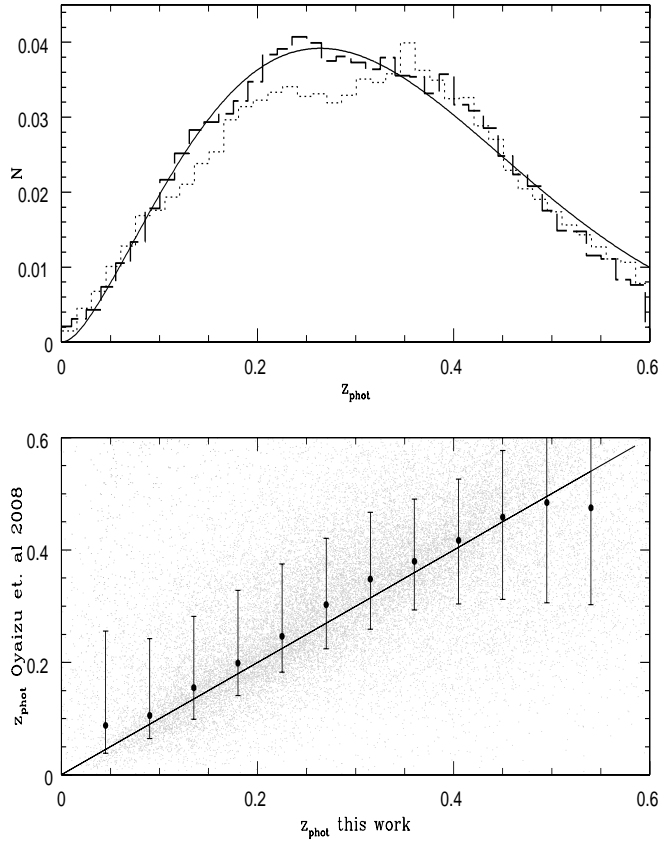}
\caption{Top panel: distribution of photometric redshifts N(z) for our calculation (short dashed) and
Oyaizu et al. 2008 (dotted) estimates. The solid curve corresponds to the expected distribution 
given by the theoretical calculation of Blanton et al. 2003. Bottom panel: the $\zphot$ redshift relation 
for Oyaizu et al. (2008) estimates and our estimates. The dashed line corresponds to the median value  and 
error bars correspond to the $10$ and $90$ percentiles associated to the median.} 
\label{f2}
\end{figure*}

\begin{figure*}
\leavevmode \epsfysize=8cm \epsfxsize=12cm \epsfbox{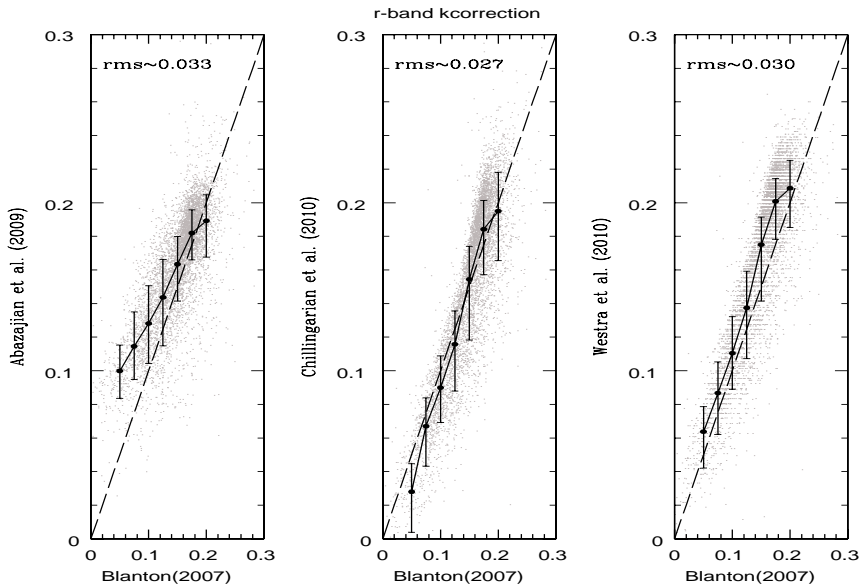}
\caption{k-correction estimated by different authors. The solid line corresponds to the median value, 
dashed line show the one to one relation and error bars correspond to the $10$ and $90$ percentiles associated to the median. 
The large scatter between different methods can be appreciated.}
\label{f3}
\end{figure*}

\begin{figure*}
\leavevmode \epsfysize=15cm \epsfbox{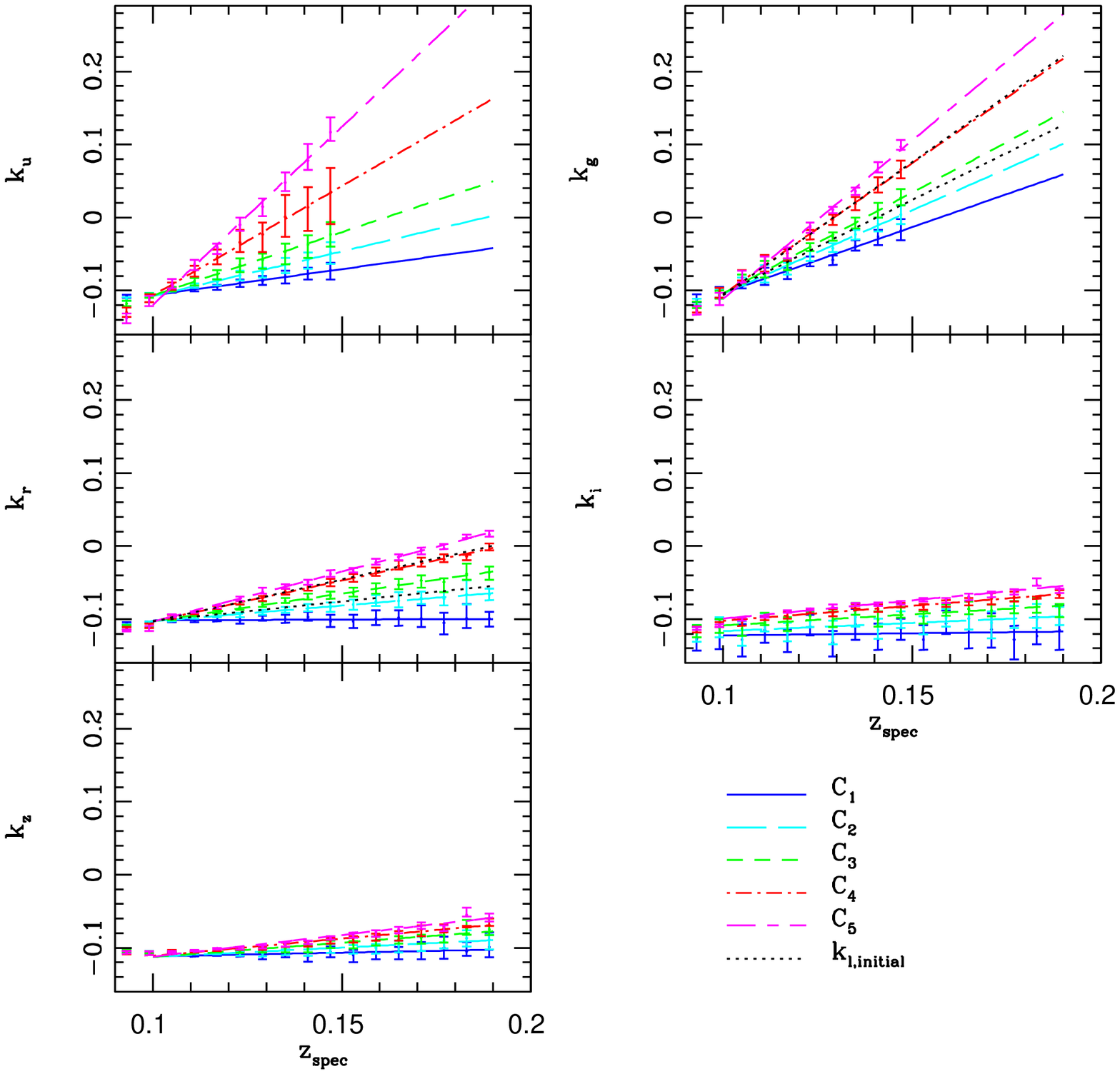}
\caption{Median K-correction estimates for the 5 SDSS photometric bands
u, g, r, i and z.  
The lines represent the fit to the median for 5 cuts in reference-colour index $(g-r)$. 
The error bars corresponds 25 and 75 percentiles from the median. 
The dotted lines in $k_g$ and $k_r$ represent the initial k values for the iteration (see text).}
\label{f4}
\end{figure*}


\subsection{Method}

We used Sz1 sample to compute photometric redshifts with the ANNz software package 
(\citet{annz}), which uses an Artificial Neural Network (ANN) to parametrize 
the relation between redshift and photometric parameters. 
ANNz is based on a "multilayer perceptron"(MLP) algorithm, where the nodes 
are disposed in layers, and the nodes in a given layer are connected to all 
the nodes in adjacent layers. The ANN topology adopted in ANNz can be 
described as $N_{in}$:$N_1$:$N_2$:...:$N_{out}$, where $N_{in}$ and $N_{out}$ 
are, respectively, the number of input and output parameters, whereas $N_i$ 
is the number of nodes in the $i-$th intermediate layer \citep[e.g.,][]{bishop}. 
The first layer contains the inputs, which in our application are photometric 
parameters. The final layer contains the outputs, in this case the photometric redshift
($\zphot$).

\begin{figure*}
\leavevmode \epsfysize=9cm \epsfbox{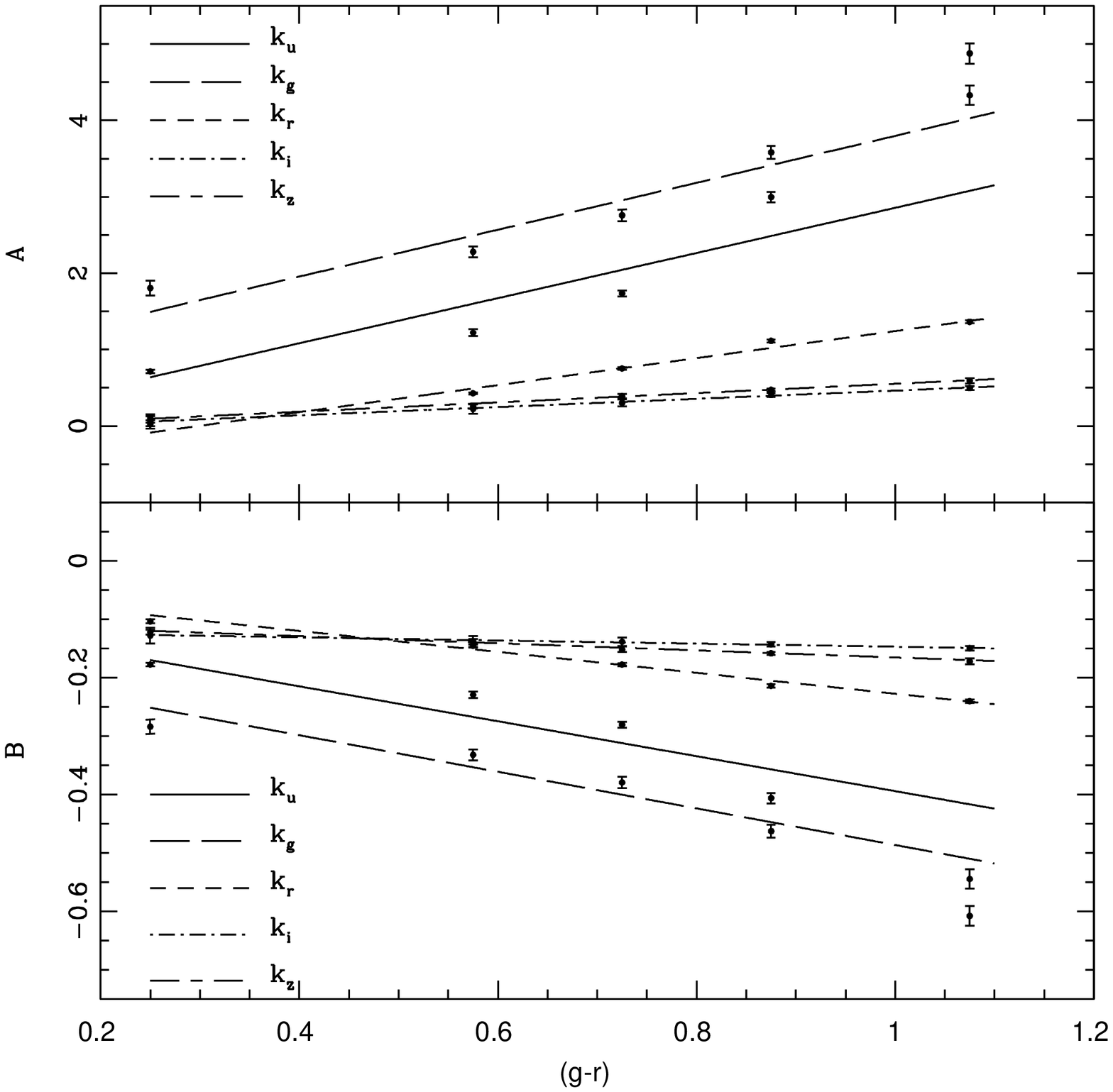}
\caption{A and B coefficients for the K correction fits of the form $k=Az+B$ 
where $A=a_A(g-r)+b_A$ and $B=a_B(g-r)+b_B$}
\label{f5}
\end{figure*}

The free parameters of ANNz are the ``weights'' between the nodes, and are
obtained by ``training'' the ANN with a training set consisting of galaxies
with spectroscopic redshifts. The selected training set must be large enough 
and representative of the target population, to assure a reliable mapping of 
the input into the output. Also, it must contain the same set of input 
parameters than the target sample, for which we want to estimate the
photometric redshifts.\\

ANNz can be trained with different sets of input parameters in order to 
improve the photometric redshift accuracy. We have analysed several sets: 
i) magnitudes in the five SDSS bands, ii) colours and 
concentration index, iii) colours and Petrosian radii, and 
iv) SDSS-DR7 magnitudes in the five bands, plus
concentration indices and Petrosian radii in $g$ and $r$-bands.
We found that the latter set is the best choice for redshift estimation. 
The use of concentration indices helps to break the degeneracies in the 
redshift-colour relation. This occurs due to the good correlation between 
the colour of a galaxy with its concentration indices and Petrosian radii.
The resulting ANNz architecture adopted here is $9:14:14:14:1$.\\

\begin{figure*}
\leavevmode \epsfysize=15cm \epsfbox{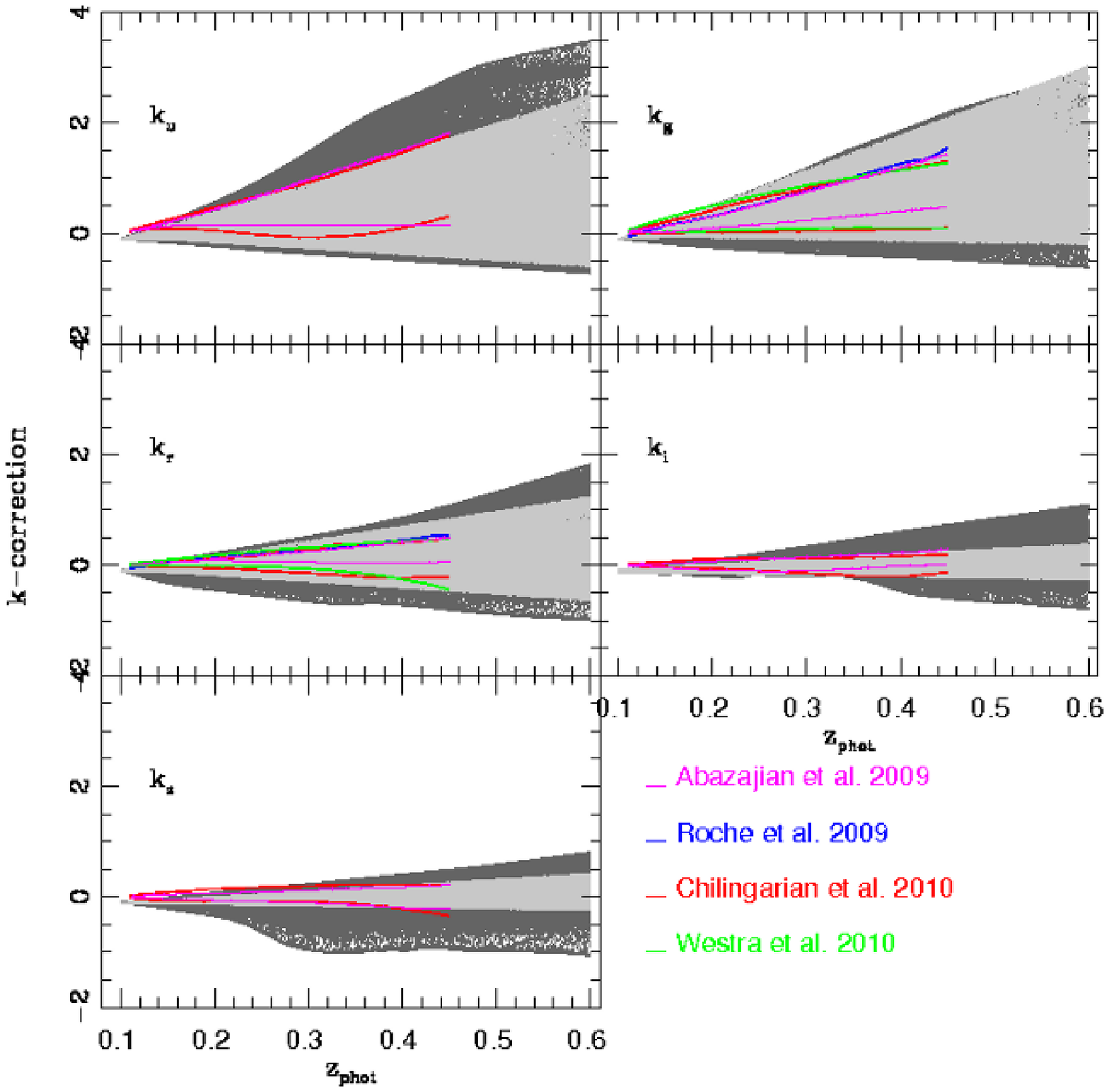}
\caption{k-correction obtained for a random sample of SDSS-DR7 galaxies. Our estimates are shown in 
grey and in dark-grey we provide the results using Blanton´s 2003 code. 
Blue solid lines are k-corrections obtained for early-type galaxies presented in \citet{roche}, 
red solid lines k-correction computed using "K-correction calculator" (\citet{chilinga}), 
magenta solid lines corresponds to \citet{dr7} estimates and in green solid lines \citet{westra} k-corrections.}

\label{f6}
\end{figure*}

\begin{figure*}
\begin{picture}(450,240)
\put(0,0){\psfig{file=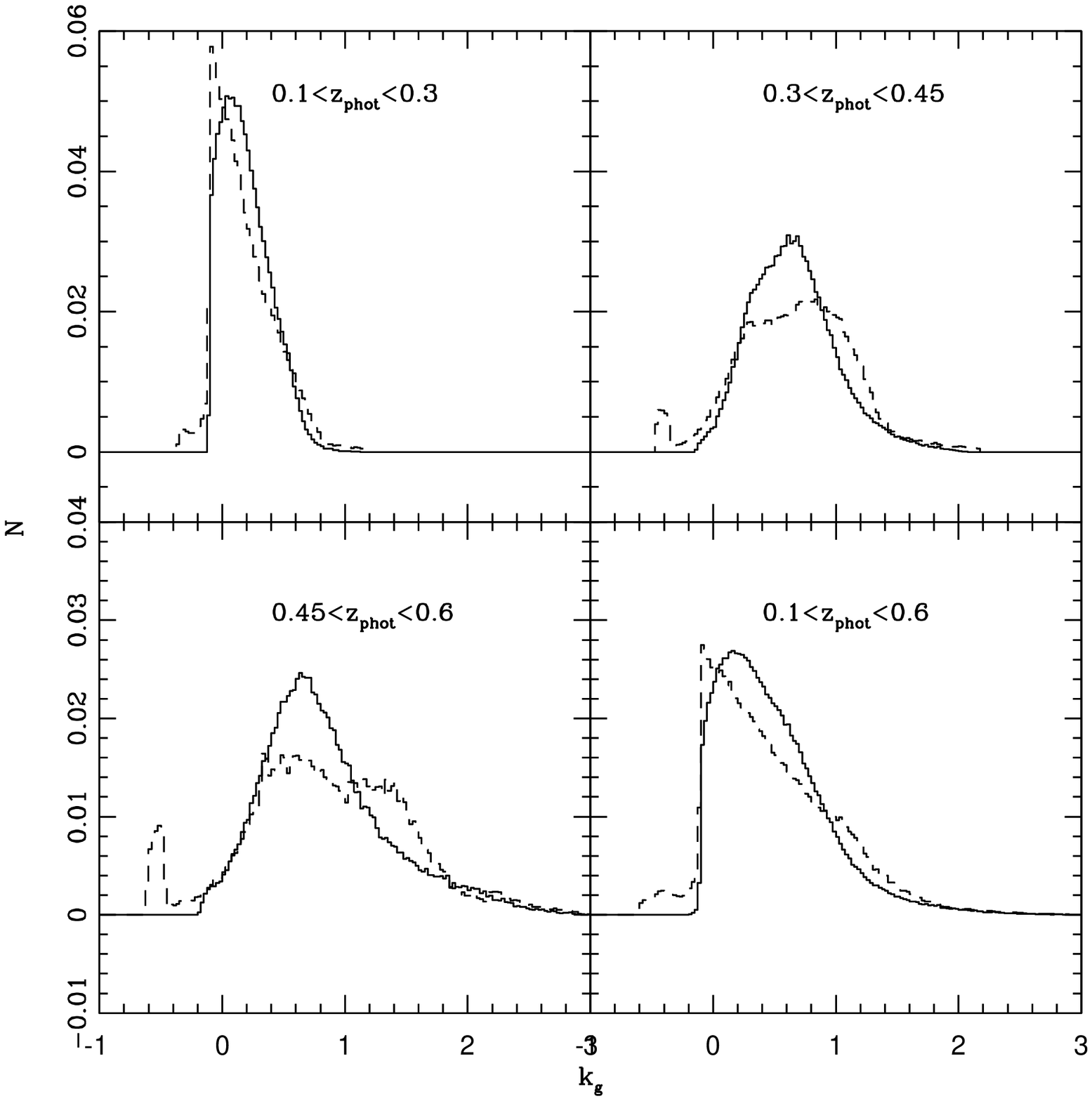,width=8cm}}
\put(240,0){\psfig{file=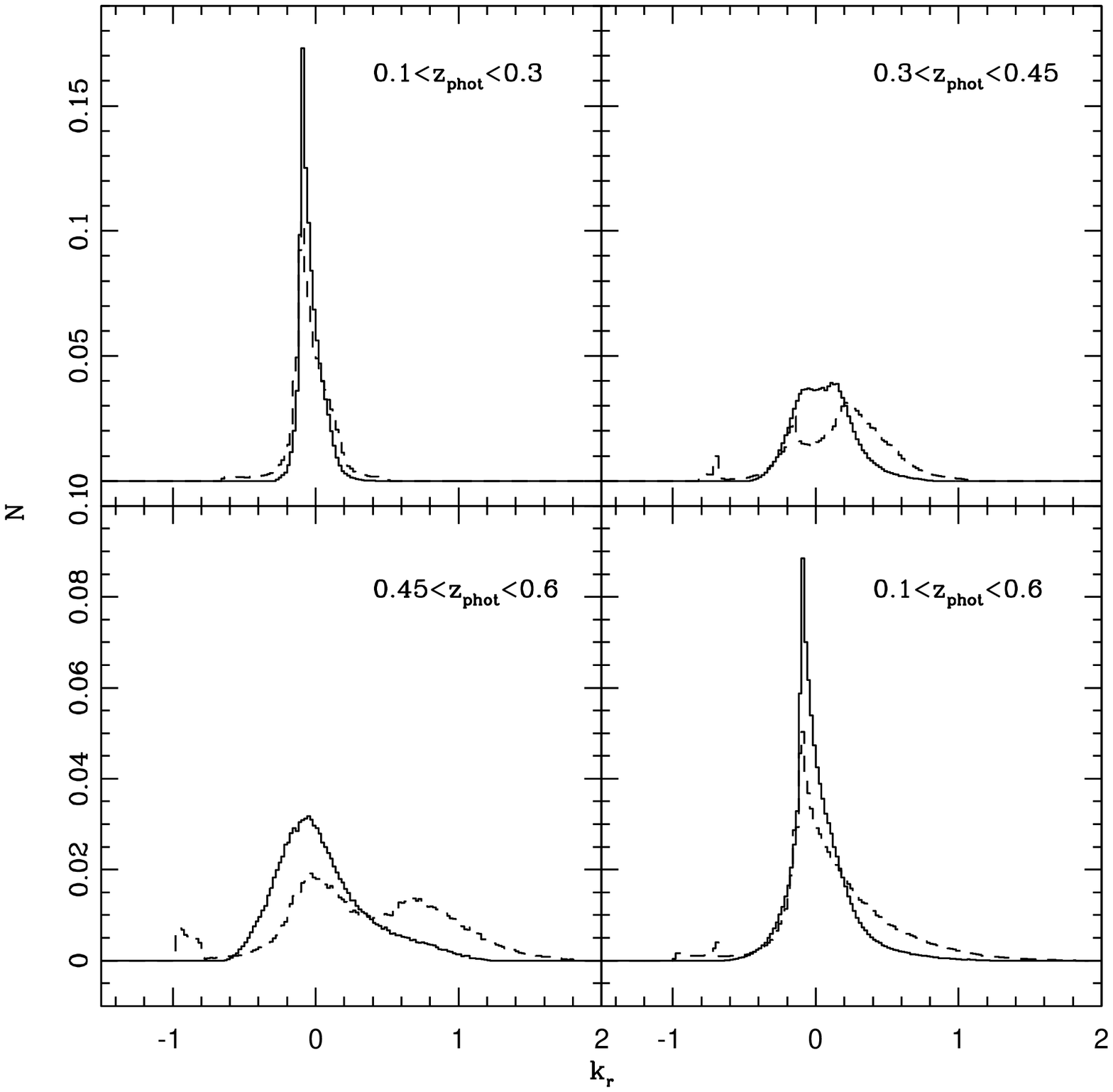,width=8cm}}
\end{picture}
\caption{$k_g$ and $k_r$ obtained for Sk2 sample in a 4 different redshift 
cuts. Our estimates are shown in solid lines and in dashed histogram we provide the results using 
k\texttt{-correct\_v4.2}}
\label{f7}
\end{figure*}


\subsection{Results}

Our results of the ANNz training are shown in Figure \ref{f1}.
The left panel of Figure \ref{f1} shows the spectroscopic ($\zspec$) vs.  
photometric ($\zphot$) redshift relation for the testing sample Sz2. As explained 
in Section 2, galaxies in the training set were excluded from the sample in order 
to avoid bias in this comparison. This figure also shows the one-to-one relation as a solid 
line, the median value of the photometric redshift estimated at a given spectroscopic redshift interval 
(dashed line) and $10$ and $90$ percentiles (as error bars) of the scatter plot. The small dispersion seen in 
this figure corresponds to  $rms \sim 0.0227$. For the MGS of the SDSS-EDR \citet{annz} obtained $rms \sim 0.0229$ 
and for a sample comprising galaxies from SDSS-MGS, SDSS-LRG, CNOC2, CFRS, DEEP2 DEEP2, TKRS and 2SLAQ surveys, \citet{zp} 
obtained $rms \sim 0.03$. The right panel of Figure 
\ref{f1} shows the systematic differences between $\zphot$ and $\zspec$, 
$z_{bias}$ (higher sub-panel), and the rms dispersion $\sigma$ (lower sub-panel)
as a function  of redshift. Both, the bias and dispersion, shown  
little dependence with the spectroscopic redshift.

For comparison with other estimates, for LRGs, \citet{abdalla} and \citet{coll2007} used the 
ANNZ code to compute a refined star/galaxy probability based on a range of photometric parameters. 
The photometric redshift rms deviation is 0.049 when averaged for all galaxies, and $0.040$ for a bright 
sub-sample with $i<19.0$ in the redshift range $\zphot=0.4$ to $0.7$.

In Figure \ref{f2} (top panel) we show the distribution of photometric redshifts $N(z)$ for Sz3 sample. 
In short dashed lines we display $N(z)$ for our calculation, and in dotted 
lines Oyaizu et al. (2008) estimates. The solid curve corresponds to the 
expected distribution given by the theoretical calculation of Blanton et al. 
(2003), assuming an universal luminosity function with parameters extracted
from SDSS data. For $\zphot < 0.1$ the observed distributions are similar to 
the expected prediction. For photometric redshifts between $\zphot \sim 0.1$ 
and $\zphot \sim 0.35$ our estimates are in better agreement with the 
theoretical curve, while Oyaizu et al. (2008) results show a $\sim 20\%$ deficit. 
We note that this range is very important because SDSS-DR7 data has 
reliable completeness up to $\zphot<0.35$, since beyond this redshift the galaxy 
density drops significantly.
For $\zphot \sim 0.4$ the observed distributions tend to overestimate the 
theoretical curve, an effect that could be associated to the Balmer break 
shifting between the $g$ and $r$ filters, difficulting the redshift estimates 
(\citet{buda01}). The bottom panel of  Figure \ref{f2} show the comparison between 
\citet{zp} $\zphot$ and our estimates. 

In Figure \ref{f2} (bottom panel) shows the $\zphot$ redshift relation 
for Oyaizu et al. (2008) estimates and our estimates. 

In \citet{ana} it is estimated the maximum redshift ($z_{max}$) beyond which compact galaxies are 
confused with the image PSF. For SDSS, a galaxy with $M_r=-23$ has a $z_{max}$ of 0.63 (see \citet{ana} Table 1). 
For this reason, we recommend using galaxies with $\zphot \leq 0.6$. 

\section{K corrections}

k-corrections allow to transform the observed magnitudes at a redshift $z$, into standard 
luminosities at some reference frame. 
This correction depends on the filter that was used for the observations, 
the rest-frame standard, the shape of the SED of the galaxy, and the redshift (see \cite{hogg01}).

In the previous sections we have used Artificial Neural Network as a tool to perform 
non-linear fitting to obtain photometric redshifts. An analysis 
based on a neural network approximation is the natural trend to obtain k-corrections. 
However these techniques needs a large and representative amount of prior information 
in the form of a training set.\\
The problem with the former is the lack of confidence on individual k-corrections estimates. 
This is shown in Figure \ref{f3} where we compare r-band k-corrections estimated by different authors 
for 6113 galaxies, selected randomly from MGS, in the narrow redshift range 0.14$<$z$<$0.16. 
For \citet{blanton07} we use \texttt{k-correct\_v4.2}, for \citet{chilinga} we employed the on-line 
available code \footnote{http://kcor.sai.msu.ru/getthecode/}, for \citet{westra} we used the 
on-line empirical k-correction calculator \footnote{http://tdc-www.cfa.harvard.edu/instruments/hectospec/progs/EOK/} 
and using models and (g-r) parameter and we also 
compare with k-corrections downloaded from the SDSS \texttt{Photoz} table (\citet{dr7}).

It is clear that individual estimates have a large uncertainty as indicated by the large
scatter between different methods, whereas the mean trends can be trusted. 

Since the development of the public code k-correct\footnote{http://howdy.physics.nyu.edu/index.php/Kcorrect}
 (\citet{blanton03}) many authors have used this method to calculate k-corrections. 
This code is based on a mathematical algorithm, namely a non-negative 
matrix factorization, which creates model based template sets. 
The set of templates is reduced to a basis of the five SDSS pass bands, which are used to interpret 
the SED of the galaxy in terms of stellar populations. Then, linear combinations of these templates 
are used to fit spectral energy distributions to broadband photometric observations for each 
galaxy and so k-corrections are obtained.

The disadvantage of this method is the use of a limited number of 
spectra that decreases with distance. The intrinsic colour of a galaxy is related to its SED, 
therefore this property can be used to compute k-corrections at high redshifts, 
where the Blanton's k-correction technique is more uncertain.

In this paper we use the public available code described 
in \citet{blanton07} (\texttt{k-correct\_v4.2}) as a tool for calibrate our k-corrections.


\subsection{Model for k-corrections}

The use of galaxy colours to describe galaxy populations presents the advantage that colours are 
easily quantifiable rather than morphological types. In this work we analyse a relation between 
k-corrections and the reference-frame galaxy $(g-r)$ colour index.

As described in Section 2, Sk1 sample comprises MGS up to $z\sim 0.2$. Since the motivation of 
this work is to derive k-corrections at higher redshifts, we have calculated k-corrections for 
the Sk1 sample using the \texttt{k-correct\_v4.2} code, analysed their dependence on the galaxy reference 
frame $(g-r)$ colour and extrapolate the results obtained to higher redshifts. 
Following \citet{blanton07}, k-corrections were calculated on the five SDSS photometric 
bands shifted to $z_{spec}=0.1$.

We divide the spectroscopic data into five different sub-samples ($C_i$) according to the reference-frame 
$(g-r)$ colour:
\begin{flushleft}
$C_1$: $(g-r)<0.5$,\\ 
$C_2$: $0.5<(g-r)<0.65$, \\
$C_3$: $0.65<(g-r)<0.8$, \\
$C_4$: $0.8<(g-r)<0.95$,\\
$C_5$: $(g-r)>0.95$.\\
\end{flushleft}

Then we calculate median values of k-corrections per redshift interval for the different $C_i$ 
sub-samples, for the five SDSS bands. The results are shown in Figure \ref{f4} where it can be 
appreciated the smooth dependence of the median k-correction values as a function of redshift 
once the reference galaxy $(g-r)$ colour is taken into account. We consider galaxies with $z\ge 0.1$ 
and we adopt a linear relation $k=Az+B$ for each $C_i$ sub-sample. The resulting fits are also 
displayed in Figure \ref{f5} where the error bars correspond to $25$ and $75$ percentiles 
of the median values.

By inspection of this figure it can be seen that our linear model is a good representation of the 
relation between k-correction and redshift. 

The values of the parameters $A$ and $B$ are obtained by fitting a straight line through
$\chi^2$ minimization for each $C_i$ sub-sample. In Figure \ref{f5} we plot the derived values of $A$ (upper panel) and
$B$ (bottom panel) as a function of the mean $(g-r)$ values for each $C_i$ sub-sample; the error
bars correspond to the square root of the variances in the estimates of $A$ and $B$ from $\chi^2$ minimization. 
From this figure it can been seen that a linear relation also gives a fair representation of 
the trends observed. Therefore, we adopt a model where 
$A=a_A(g-r)+b_A$ and $B=a_B(g-r)+b_B$. The final calibration of Blanton's k-corrections is 
given as 
\begin{equation}
k_j=[a_{Aj}(g-r)+b_{Aj}]z+[a_{Bj}(g-r)+b_{Bj}]
\label{eq1}
\end {equation}
where $j$ represents the different $ugriz$ bands.

In Table \ref{t2} we present the  $a_{Aj}$, $b_{Aj}$, $a_{Bj}$ and $b_{Bj}$ values and the errors obtained by 
$\chi^2$ minimization.\\

\begin{table*}
\begin{minipage}{175mm}
\caption{Parameters of the $k_j=[a_{Aj}(g-r)+b_{Aj}]z+[a_{Bj}(g-r)+b_{Bj}]$ relation obtained by $\chi^2$ minimization}
  \begin{center}\begin{tabular}{@{}ccccccccc@{}}
  \hline
  \hline
Band & $a_A$ & $\sigma_{a_A}$ & $b_A$ & $\sigma_{b_A}$ & $a_B$ & $\sigma_{a_B}$ & $b_B$ & $\sigma_{b_B}$ \\
 \hline

$u$ &  $2.956$ &  $0.070$ & $-0.100$ &  $0.034$ &  $-0.299$ &  $0.009$ &  $-0.095$ &  $0.004$\\
$g$ &  $3.070$ &  $0.165$ & $0.727$  &  $0.117$ &  $-0.313$ &  $0.021$ &  $-0.173$ &  $0.015$\\
$r$ &  $1.771$ &  $0.032$ & $-0.529$ &  $0.023$ &  $-0.179$ &  $0.005$ &  $-0.048$ &  $0.003$\\
$i$ &  $0.538$ &  $0.085$ & $-0.075$ &  $0.079$ &  $-0.027$ &  $0.013$ &  $-0.120$ &  $0.012$\\ 
$z$ &  $0.610$ &  $0.045$ & $-0.064$ &  $0.034$ &  $-0.061$ &  $0.007$ &  $-0.106$ &  $0.005$\\ 
\hline
\hline
\label{t2}
\end{tabular}
\end{center}
\end{minipage}
\end{table*}

\subsection{Estimation of k-corrections} 

In order to compute k-corrections using the model described in the previous Section, 
it is necessary to obtain reference-frames $(g-r)$ colours for each galaxy, which in 
turn depends on k-corrections. 

To avoid this problem we have adopted an iterative procedure as follows: 
the iteration starts with an initial value for $k_g$ and $k_r$ according to the value 
of the concentration index of the galaxy in the r band ($c_r$). This quantity does not 
depend on k-correction and is a suitable indicator of galaxy morphological type bimodality: 
early-type galaxies have higher $c_r$ than later types (\citet{strateva}, \citet{kauf2003a}, 
\citet{kauf2003a}, \citet{mateus06}). 

According to \citet{strateva}, $c_r>2.55$  values correspond to early-type galaxies, 
and late-type galaxies have $c_r <2.55$. This bimodality can also be seen in the colour 
distribution of galaxies. In particular, the distribution of $(g-r)$ colours has two well-defined 
peaks: one at $(g-r)= 0.60$ and other at $(g-r) = 0.95$, corresponding to the late and early-type 
components, respectively. Taken these properties into account, we assign $(g-r)_{initial} = 0.60$ 
if $c_r <2.55$  and $(g-r)_{initial} = 0.95$ if $c_r >2.55$ so that the initial $k_g$ and $k_r$ 
values are: $$k_{l,initial}=[a_{Al}(g-r)_{initial}+b_{Al}]z+[a_{Bl}(g-r)_{initial}+b_{Bl}],$$ 
where $l$ refers to the $g$ and $r$ bands (see the dotted lines Figure \ref{f4})

Once these initial values are fixed, we iterate in equation \ref{eq1} to obtain $k_g$ and $k_r$ for each galaxy. 
After each iteration, we check that the colours are within an acceptable range, $0<(g-r)<1.8$ 
in order to avoid either too red or too blue colours. In these cases the iteration starts again 
with a new $(g-r)_{initial}$ value selected at random from a Gaussian distribution 
(within $1 \sigma$) that fits either the blue or the red peak of the reference frame 
colour distribution.

The iterations stop when the difference in both $k_g$ and $k_r$ between two 
consecutive steps is less than $0.001$. This procedure converges in less than 15 iterations.

From the finally obtained $k_g$ and $k_r$ values we calculate the reference frame $(g-r)$ 
colours, which allows to compute k-corrections in the other bands using equation \ref{eq1}. 

We notice that there is a small percentage (less than $0.4\%$) of galaxies 
for which our algorithm does not converge. Howerver, the main sources for this lack of convergence 
are large magnitude uncertainties and unreliable observed colours($(g-r)_{obs}>3$). 
The last colours constraints are helpful on removing stars with unusual colours, without 
discarding real galaxies (\citet{lopes}, \citet{padmanabhan}). \citet{coll2007} point out that the 
stellar contamination may still be present.

\begin{figure*}
\leavevmode \epsfysize=9cm \epsfbox{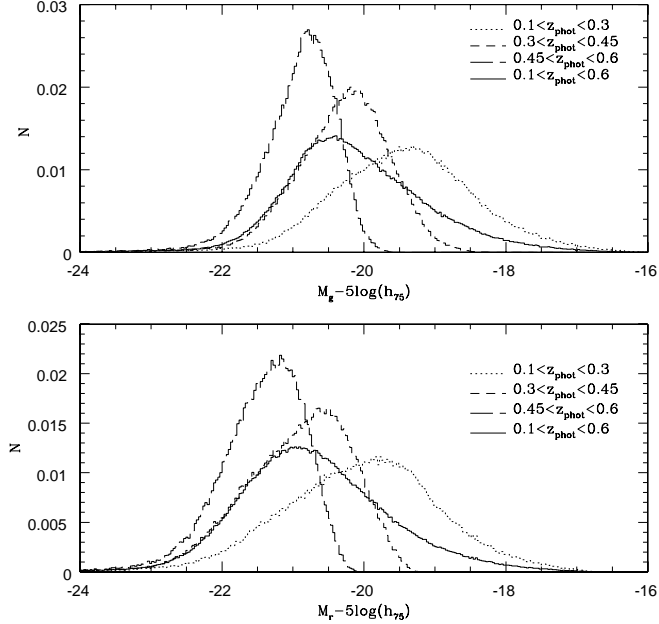}
\caption{Luminosity distribution for galaxies brighter 
than $M_r-5log(h_{75})=-21.5$ and for 4 different redshifts cuts.}
\label{f8}
\end{figure*}

\begin{figure*}
\begin{picture}(450,240)
\put(0,0){\psfig{file=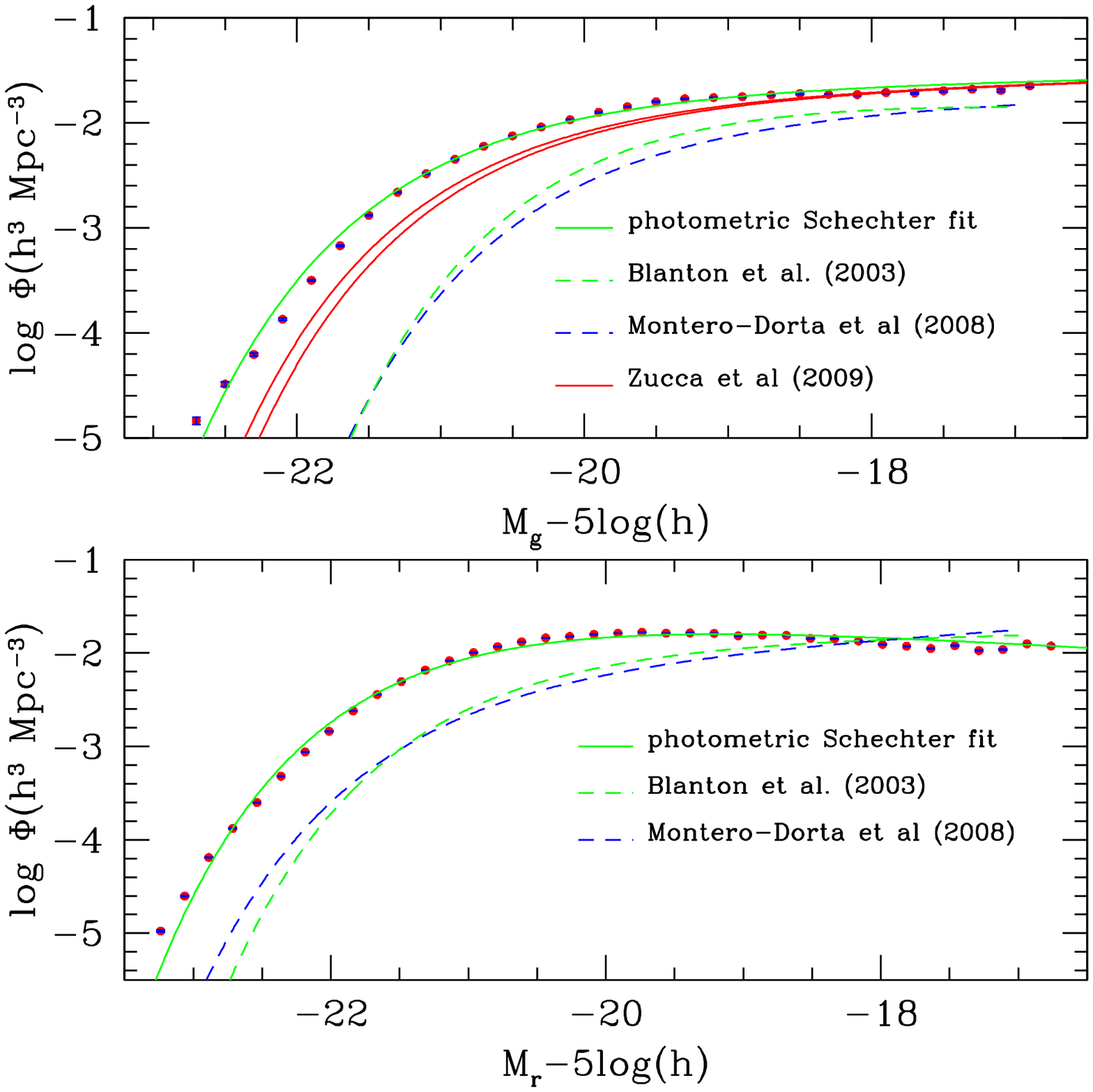,width=8cm}}
\put(240,0){\psfig{file=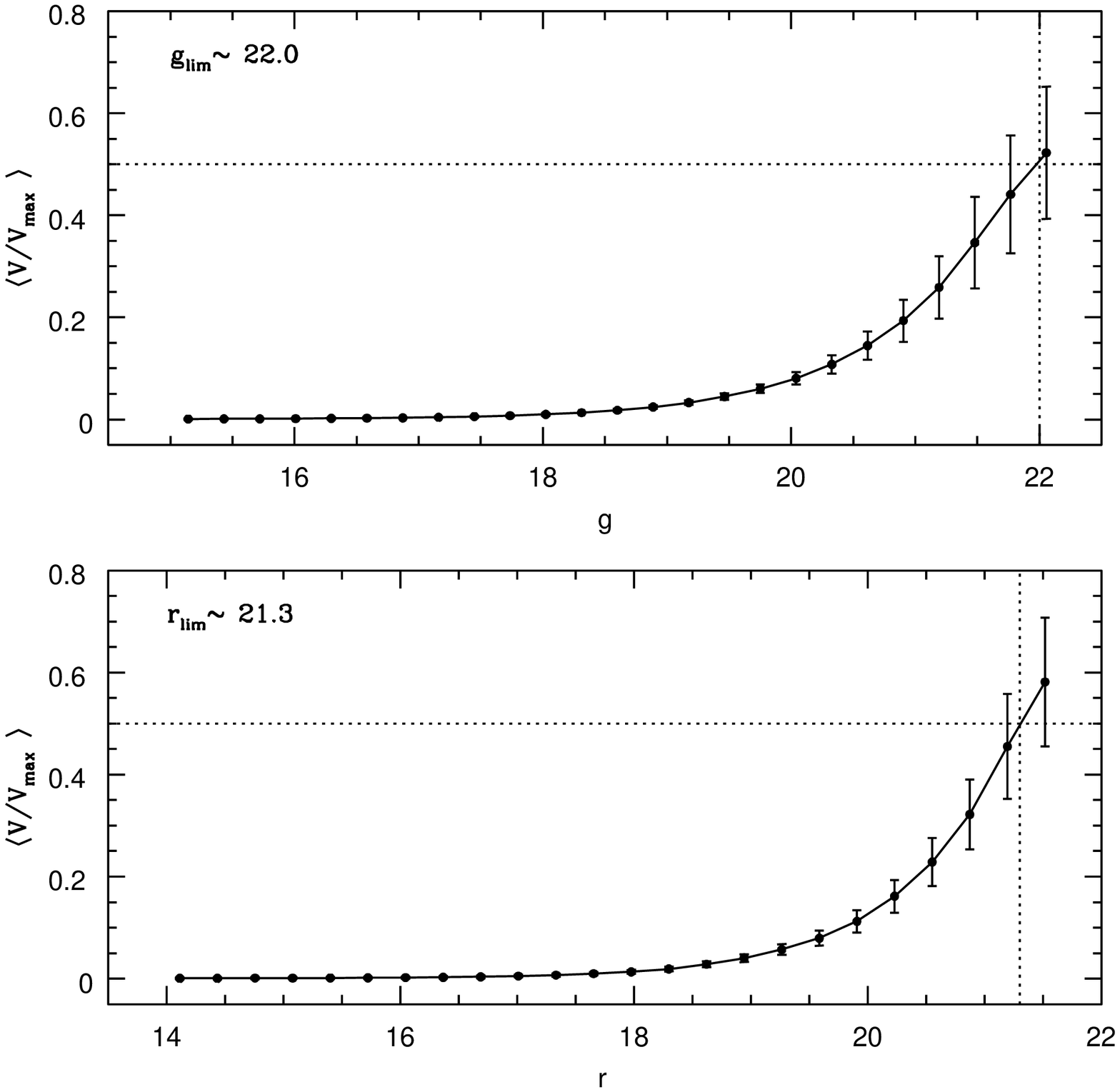,width=8cm}}
\end{picture}
\caption{Left panel: Luminosity function in $g$ and $r$ bands. The dashed lines show the Schecter fit 
for MGS of \citet{blanton03} and \citet{maglim} (green and blue respectively). 
The green solid line is the result from the $1/Vmax$ method for our photometric sample. 
The red solid lines are the result from the $STY$ method from two different redshift bins 
($0.1<z<0.35$ and $0.35<z<0.55$) from \citet{zucca}. The error bars (blue colours) represent 
$1\sigma$ uncertainty calculated using a bootstrapping technique. Right panels: $V/Vmax$ test. The error bars represent $1\sigma$ deviation from the median. 
The horizontal dotted lines at $V/Vmax=0.5$ correspond to the $g$ and $r$ band completeness.}
\label{f9}
\end{figure*}

\begin{figure*}
\leavevmode \epsfysize=9cm \epsfbox{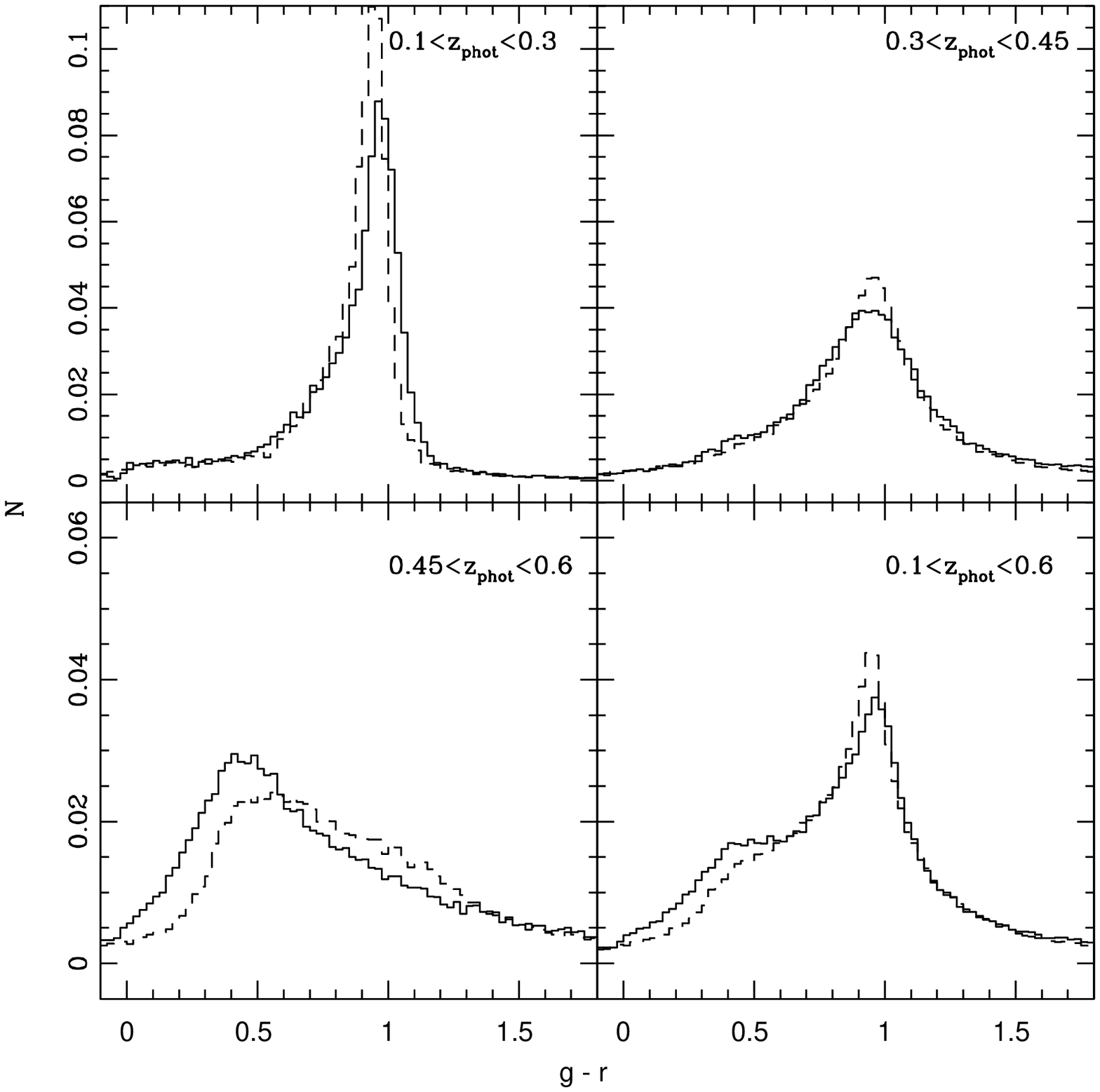}
\caption{Derived rest frame $(g-r)$ colour distribution for galaxies brighter 
than $M_r-5log(h_{75})=-21.5$ and for 4 different redshifts cuts in Sk2 sample. The dashed histogram corresponds to Blanton's 2003 
results and in solid lines our estimates. Relative excess of red galaxies in Blanton´s results is clearly seen 
in comparison to our colour estimates}
\label{f10}
\end{figure*}

\subsection{Results} 

We have compared the k-corrections obtained in this work with the results from
\texttt{k-correct\_v4.2} using the Sk2 sample. As explained in Section 2, we have 
restricted our analysis to galaxies with $r<21.5$. 

In Figure \ref{f6} we plot both estimations against redshift for the five 
SDSS-DR7 bands. In dark-grey we show \texttt{k-correct\_v4.2} results whereas
light-grey points are our estimates. It can bee seen 
in all cases, that our k-correction shows the same trend but with a lower spread than \citet{blanton07} 
results. This lower spread is particularly noticeable in the $i$ and $z$ bands. 
However, our estimations are in agreement with \citet{chilinga} k-corrections in these bands. 
In Figure \ref{f6} we also compare our k-corrections with those obtained for $E/S0$ galaxies by \citet{roche}. 
We find that the mean k-correction for the early type galaxies lies within our estimates for 
red galaxies ($C_5$ sample), where k-correction values are higher. 
We also perform a comparison with k-corrections obtained using the ``K-correction 
calculator'' (\citet{chilinga}), the ``on-line empirical k-correction calculator'' \citep{westra} 
and k-corrections from \texttt{Photoz} table \citep{dr7}. We compute the k-corrections in 
all bands at extreme red and blue colours, finding that these estimates are within our 
range of calibrated values. We notice a difference in the $k_u$ for blue galaxies at higher redshift 
with respect to \citet{chilinga}. We argue that this can be originated in the fact that these 
authors calibrate $k_u$ with the $u-r$ colour and that 
the $u$ filter has a natural red leak that cases abnormal $u-r$ 
colours. This effect can propagate to the k-corrections, generating higher values, particularly at large redshifts.

In Figure \ref{f7} we plot the distributions of $k_g$ and $k_r$ in four 
different redshift intervals. 
Our estimates are shown as solid lines, and the dashed histogram is 
\texttt{k-correct\_v4.2} code \citep{blanton07} 
results. For galaxies in the redshift range $0.1<\zphot<0.3$, both distributions are similar. 
As redshift increases, our distributions remain always unimodal,
with a well defined mean. 
On the other hand, Blanton's k-correction distribution shows 3 different maxima, one centered approximately in 
our distribution, and the others in the extremes. 
This behavior is probably indicative of 
template mismatch. At redshifts $0.1<\zphot<0.6$, both $k_r$ distributions approximately match each 
other, whereas $g$-band k-corrections from \texttt{k-correct\_v4.2} distribution is lopsided to negative values.

\subsection{Luminosities and colours}

We have estimated the absolute magnitude of galaxies in the Sk2 sample in the $g$ and $r$ 
bands using our k-corrections. The absolute magnitude of a galaxy in a given band is related 
to its apparent magnitude by: $$M_l= l - 5log_{10} (DL(z))-25 - k_l$$ where $l$ refers to 
the $g$ and $r$ bands, and $DL(z)$ is the luminosity distance (which depends on the cosmological 
parameters adopted). In Figure \ref{f8}, we show the distribution of k-corrected absolute magnitudes 
in four different increasing redshift intervals (key in Figure). 
The shape of these distributions is very similar in both $g$ and $r$ SDSS bands: a bell-shaped 
distribution skewed to fainter magnitudes. Notice that the mean of the distributions move towards 
fainter magnitudes as the redshift range decreases, while the bright tail of the distribution 
remains approximately fixed at $M_g \sim -22.5$ and $M_r \sim -23$.

For Sk2 photometric sample we computed the luminosity function in the $g$ and $r$ bands using the $1/Vmax$ method 
considering the incompleteness with a $V/Vmax$ test (\citet{schmidt}). 
This methods takes into account the volume of the survey enclosed by the galaxy redshift and the difference between the maximum and minimum 
volumes within which it can be observed. 

In Figure \ref{f9} we show the luminosity function and $V/Vmax$ test for the redshift range $0.1<\zphot<0.6$. 
The dashed lines show the Schechter fit for MGS of \citet{blanton03} and \citet{maglim}. 
The green solid line is the resulting Schechter fit for Sk2 sample. 
We compared our result in $g$ band ($4686\AA$) with two different redshift bins from \citet{zucca}. 
These authors have studied the evolution in the $B$ band ($4459.7\AA$) luminosity function 
to redshift $z\sim 1$ in the $zCOSMOS$ from the $STY$ method. 

Our best Schechter fit corresponds to $M^{\star}\sim -20.469 \pm 0.0535$ , $\Phi^{\star}\sim 0.0224 \pm 0.0089$ 
and $\alpha \sim -1.065  \pm 0.0286$ for $g$ band, and $M^{\star}\sim -20.821 \pm 0.0966$, $\Phi^{\star}\sim 0.030 \pm 0.0029$ 
and $\alpha \sim -0.78 \pm 0.0298$ for $r$ band. Where the errors had been obtained through bootstrapping. 
We notice that the luminosity functions at $z \sim 0.5$ from our work  and that 
of \citet{zucca} are consistent taking into account the $(g-B)$ values of typical Sbc galaxies at this redshift 
\citep{fuku95}.

In Figure \ref{f10} we compare the $(g-r)$ colour distribution of galaxies brighter 
than $M_r=-21.5$ in four redshift intervals. The dashed histogram corresponds to 
\texttt{k-correct\_v4.2} code results, and in solid lines we show our estimates. 
For galaxies with $0.1<\zphot <0.3$ both distributions are similar, 
exhibiting a prominent red peak corresponding to early-type galaxies. As 
redshift increases, the distributions are shifted bluewards, and we notice in 
the redshift range $0.45<\zphot<0.6$ an excess of red galaxies in \citet{blanton07} 
results in comparison to ours. 
This trend can also be seen in the full redshift range (Figure \ref{f10} bottom right panel), 
where the galaxy bimodality can also be appreciated. 

\begin{figure*}
\leavevmode \epsfysize=15cm \epsfbox{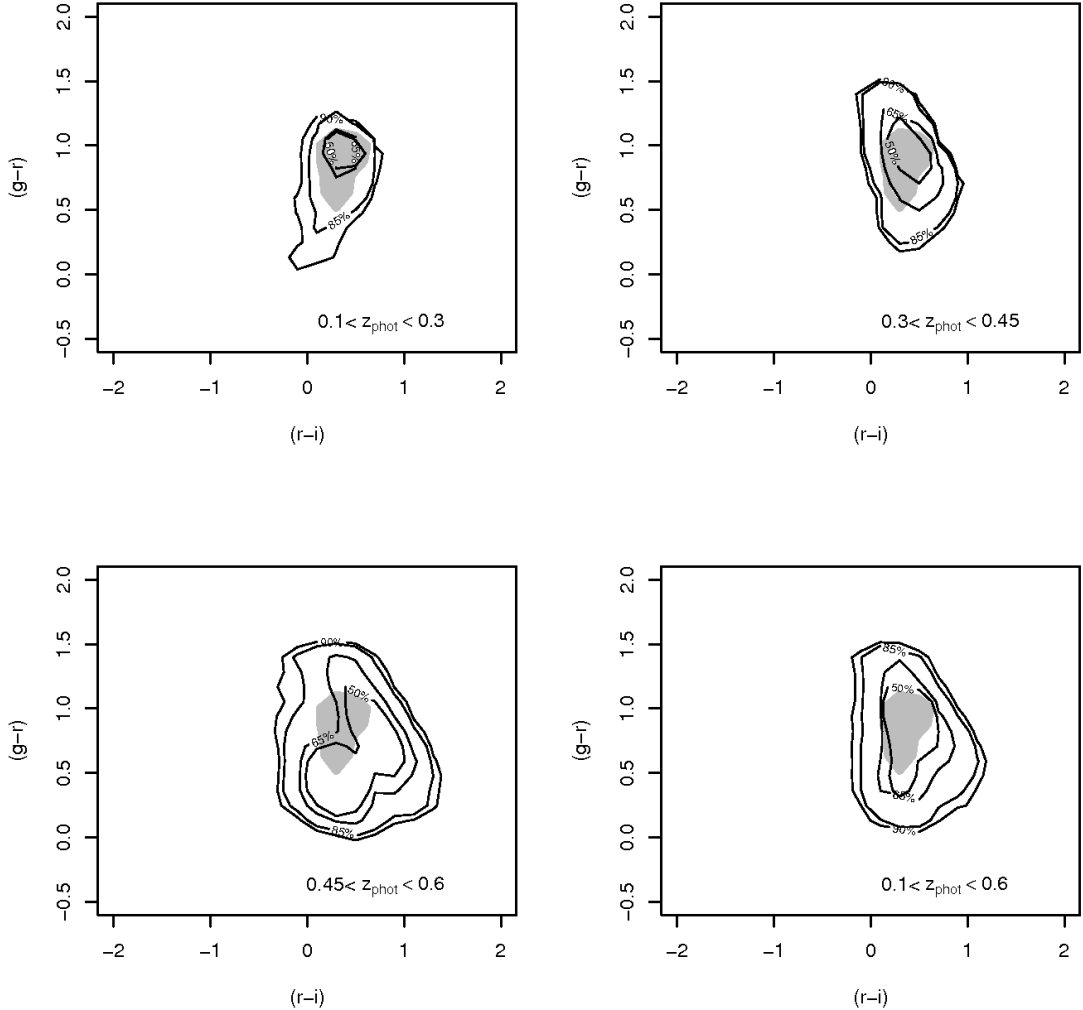}
\caption{ $(g-r)$ vs $(r-i)$. colour-colour diagram for galaxies brighter than $M_r=-21.5$ 
in four different redshift intervals for Sk2 sample. The contours enclose $50\%$, $65\%$ $85\%$, and $90\%$ 
of the photometric sample. The shaded region represents the colour-colour diagram for $90\%$ of MGS.} 
\label{f11}
\end{figure*}

\begin{figure*}
\leavevmode \epsfysize=15cm \epsfbox{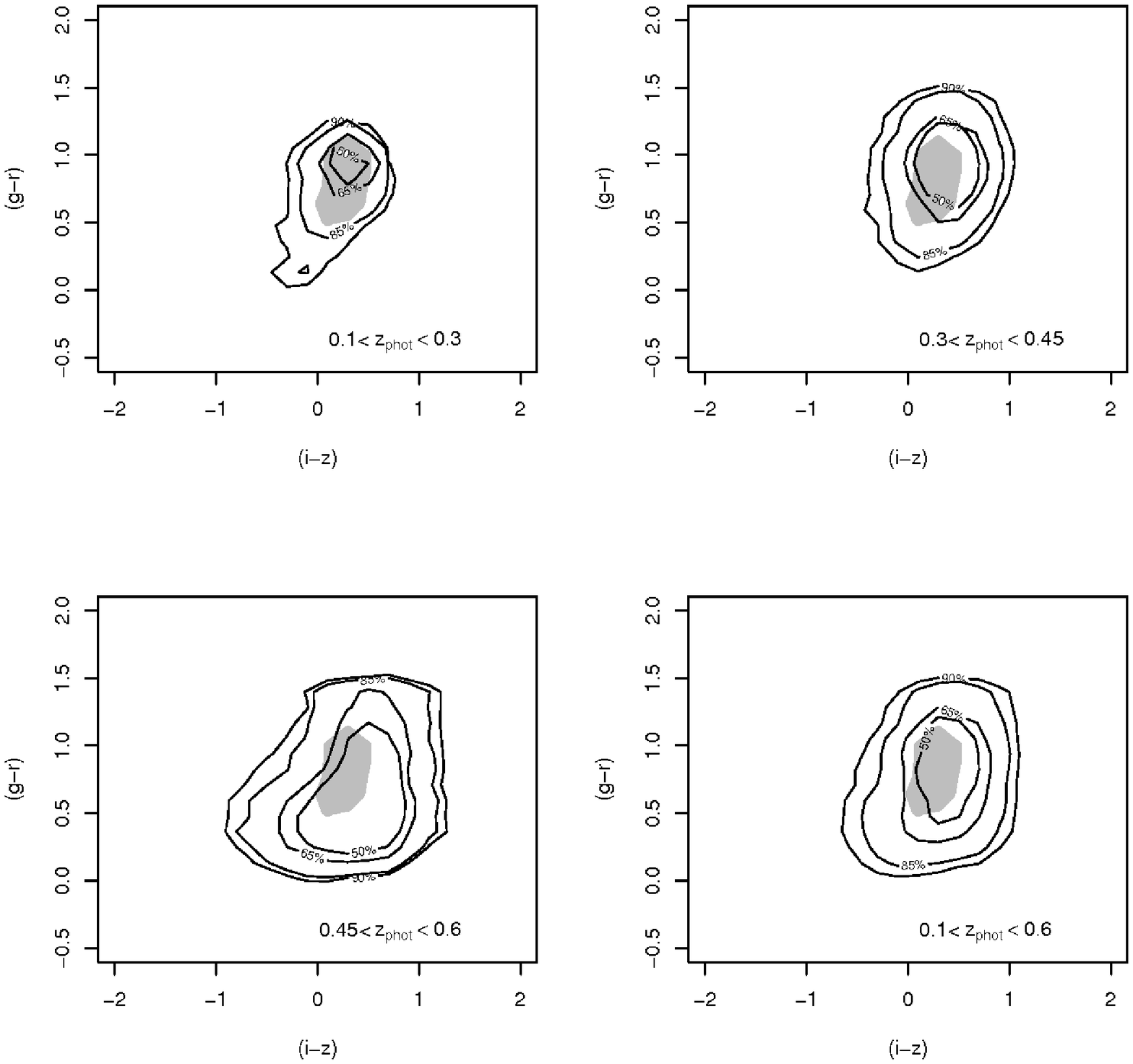}
\caption{$(g-r)$ vs $(i-z)$ colour-colour diagram for galaxies brighter than $M_r=-21.5$ 
in four different redshift intervals for Sk2 sample. The contours enclose $50\%$, $65\%$ $85\%$, and $90\%$ 
of the photometric sample. The shaded region represents the colour-colour diagram for $90\%$ of MGS.} 
\label{f12}
\end{figure*}

In Figures \ref{f11} and \ref{f12} we show the colour-colour diagrams  
$(g-r)$ vs $(r-i)$ and $(g-r)$ vs $(i-z)$ for galaxies brighter than 
$M_r=-21.5$ in four different redshift intervals. The contours enclose 
$50\%$, $65\%$ $85\%$, and $90\%$ of the galaxies in Sk2 sample. In all cases 
the shaded region represents the corresponding colour-colour diagram for 
$90\%$ of MGS, which have spectroscopic redshifts. 
The left upper panel ($0.1<\zphot<0.2$) shows that contours calculated for 
the photometric sample matches the shaded region. As the redshift increases, 
the contours in the colour-colour diagrams expand with respect to the spectroscopic 
sample, but both have approximately the same center. 
The expansion of the contours for the photometric results could be due to 
uncertainties in the determination of photometric redshifts and k-corrections, 
as well as they may reflect galaxy evolution. The colours show a decreasing trend 
for the red population as well as a constant shift to bluer colours (\citet{ana}). 
This is particularly important for the redshift range $0.45<\zphot<0.6$.

\section{Conclusions}

In this work we present a new set of photometric redshift ($\zphot$) 
and k-correction estimations for the SDSS-DR7 photometric catalogue available on the World Wide 
Web.
In order to calculate $\zphot$, artificial neural networks were applied 
using the public code ANNz. The improvements in the SDSS-DR7 photometric 
redshift estimation are:\\
1) We added the concentration index and the Petrossian radii in $g$ and $r$ bands to the usual five 
magnitudes used in previous similar works. These additional inputs improve 
$\zphot$ estimations because the concentration index provides information regarding the slope of the galaxy brightness 
profile, helping us to break the degeneracies in the redshift-colour relation
due to morphology. 
The Petrossian radius is a robust measure of how shallow the
brightness profile is and contain information about the angular size, 
that is related to the distance.\\
2) The choice of different galaxy samples for the training set (MGS, LRG and AGN sample) 
provides a wide sampling of different galaxy types at various redshifts, 
allowing to improve $\zphot$ estimates.

Our $\zphot$ estimates have a $rms\simeq 0.0227$, and the resulting galaxy
distribution shows a good agreement with the theoretical distribution 
derived from the SDSS galaxy luminosity functions.

We have used the \texttt{k-correct\_v4.2} code (\citet{blanton07}) for the MGS 
and we have performed a linear fit between reference frame $(g-r)$ 
colour and redshift, extrapolating this relation at high redshifts. 
We propose an iterative procedure to estimate k-corrections from the
observed photometry in the $g$ and $r$ bands. 
Using initial values that depend on the concentration index and the observed 
colour, we obtain the 
k-correction for the other bands. Our results show that the use of this 
simple linear relation between the reference frame $(g-r)$  colour and 
redshift is as accurate as those obtained in previous work. A clear plus 
of our approach is the low computational time.
  
Our k-correction estimations do not use templates, avoiding statistical errors in the lack of homogeneity 
in spectral features, and minimizing systematical errors caused by an assumed spectral 
energy distribution (SED). This can be noticed in the smooth behavior of the distribution of k-corrections,  
even for intermediate redshifts.

The analysis of the distribution of k-corrected absolute magnitudes show that the 
shape of these distributions has a bell-shaped skewed to fainter magnitudes and 
the mean of the distributions move towards fainter magnitudes as the 
redshift range decreases.

We have computed the luminosity function in $g$ and $r$ bands trhough $1/Vmax$ method taking into account the 
incompleteness with a $V/Vmax$ test (\citet{schmidt}). We notice that the curves derived from this work and that 
of \citet{zucca} are consistent taking into account the $(g-B)$ value of a typical Sbc galaxy at 
$z \sim 0.5$ \citet{fuku95}.

The analysis of the $(g-r)$ colour distribution for galaxies brighter 
than $M_r=-21.5$ shows that the galaxies in our samples 
are shifted bluewards as redshift increases. 
This trend leads to the emergence of bimodality in the full redshift range.
From the colour-colour diagrams, we can conclude that the behavior 
of colours at low redshift is in good agreement with the trends of the spectroscopic sample. 
As redshift increases we see a broadening of the contours and an increase 
in the  blue galaxy population. However, the distribution of galaxies in the colour-colour diagram remains 
centered with respect to the spectroscopic data.

\section{Acknowledgments}

We thank the Referee for very helpful comments which that greatly improved this paper. 
We thank William Schoenell and the The SEAGal/STARLIGHT Project.
This work was supported in part by the Consejo Nacional de 
Investigaciones Cient\'ificas y T\'ecnicas de la Rep\'ublica Argentina 
(CONICET), Secretar\'\i a de Ciencia y Tecnolog\'\i a de la Universidad 
dad de C\'ordoba. Laerte Sodr\'e Jr was supported by the Brazilian agencies FAPESP and CNPq. 
Funding for the SDSS and SDSS-II has been provided by the Alfred P. Sloan Foundation, 
the Participating Institutions, the National Science Foundation, the U.S. 
Department of Energy, the National Aeronautics and Space Administration, 
the Japanese Monbukagakusho, the Max Planck Society, and the Higher Education 
Funding Council for England. The SDSS Web Site is http://www.sdss.org/. 
The SDSS is managed by the Astrophysical Research Consortium for 
the Participating Institutions. The Participating Institutions are 
the American Museum of Natural History, Astrophysical Institute 
Potsdam, University of Basel, University of Cambridge, Case 
Western Reserve University, University of Chicago, Drexel 
University, Fermilab, the Institute for Advanced Study, the Japan 
Participation Group, Johns Hopkins University, the Joint Institute 
for Nuclear Astrophysics, the Kavli Institute for Particle 
Astrophysics and Cosmology, the Korean Scientist Group, the 
Chinese Academy of Sciences (LAMOST), Los Alamos National 
Laboratory, the Max-Planck-Institute for Astronomy (MPIA), the 
Max-Planck-Institute for Astrophysics (MPA), New Mexico State 
University, Ohio State University, University of Pittsburgh, 
University of Portsmouth, Princeton University, the United States 
Naval Observatory, and the University of Washington.


\label{lastpage}

\end{document}